# Generating unfavourable VaR scenarios with patchwork copulas

Dietmar Pfeifer[1] and Olena Ragulina[2]

Carl von Ossietzky Universität Oldenburg, Germany[1] and
Taras Shevchenko National University of Kyiv, Ukraine[2]

May 5, 2021

**Abstract:** The central idea of the paper is to present a general simple patchwork construction principle for multivariate copulas that create unfavourable VaR (i.e. Value at Risk) scenarios while maintaining given marginal distributions. This is of particular interest for the construction of Internal Models in the insurance industry under Solvency II in the European Union.

**Key words:** copulas, patchwork copulas, Bernstein copulas, Monte Carlo methods

**AMS Classification:** 62H05, 62H12, 62H17, 11K45

## 1. Introduction

Reasonable VaR-estimates from original data or suitable scenarios for risk management within so-called Internal Models are – besides the banking sector under Basel III – of particular interest in the insurance industry under Solvency II (see, e.g., Cadoni (2014); Cruz (2009); Doff (2011,2014); McNeil et al. (2015); Arbenz et al. (2012) or Sandström (2011)). In this paper, we propose a simple stochastic Monte Carlo algorithm on patchwork copulas for the generation of VaR scenarios that are suitable for comparison purposes in Internal Models for the calculation of Solvency Capital Requirements (SCR), in particular for the Non-Life Module. Note that in the Standard Formula of Solvency II, there is a formula for the calculation of the non-life premium and reserve risk SCR given by the volume factor

$$\rho_{1-\alpha}(\sigma)_{\text{VaR}} = \frac{\exp\left[k_{1-\alpha} \cdot \sqrt{\ln(1+\sigma^2)}\right]}{\sqrt{1+\sigma^2}} - 1$$

applied to the volume measure (i.e. premium income) of the year considered (see e.g. Sandström (2011), relation (21.9b), p. 324; cf. also Hürliman (2009), p. 329 ff.). Here $\alpha$ denotes the risk level (i.e. 0.05% in Solvency II) and $k_{1-\alpha}$ the corresponding $1-\alpha$ quantile of the standard normal distribution. Further, $\sigma$ denotes the standard deviation of the underlying risk, i.e. the ultimate combined loss ratio which is assumed to be lognormally distributed with expectation $1 = 100\%$ (which is the limit towards certain ruin according to the law of large numbers). However, this formula is questionable from a scientific point of view (see Pfeifer and Hampel (2011)). Note also that this formula was simplified in the Commission Delegated Regulation of the EU (2015), Article 115:

$$\rho_{1-\alpha}(\sigma)_{\text{VaR}} \approx 3\sigma \text{ for } \alpha = 0.005.$$

This is a reasonable conservative approximation as long as $\rho < 0.15$, see Fig. 1.

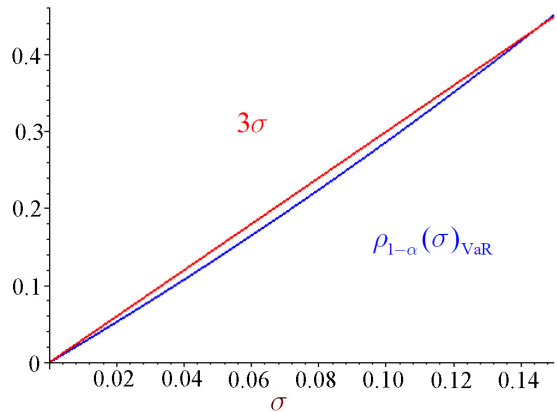

**Figure 1.** Plot of the Non-Life SCR volume factor $\rho_{1-\alpha}(\sigma)_{\text{VaR}}$ vs. its simplification $3\sigma$

Another questionable point here is the aggregation to the overall SCR from different module SCR's by correlations in Solvency II, see e.g. Sandström (2011). This has been discussed in detail e.g. in Pfeifer and Straßburger (2008) and Pfeifer (2013).

Note that the European Union (2015) concerning the implementation of Solvency II in the EU requires the consideration of risk scenarios in several Articles, in particular in Article 259 on Risk Management Systems saying that insurance and reinsurance undertakings shall, where appropriate, include performance of stress tests and scenario analyses with regard to all relevant risks faced by the undertaking, in their risk-management system. The results of such analyses also have to be reported in the ORSA (Own Risk and Solvency Assessment, see e.g. Ozdemir (2015)) as described in Article 306 of the Commission Delegated Regulation. In the light of the outlined structural problems with the standard formula above, the ORSA is probably a better instrument to rate the enterprise's risks in a more reliable way. The problem is, however, that the Commission Delegated Regulation does not make any clear statements on how such stress tests or scenario analyses have to be performed.

Article 1 of the Commission Delegated Regulation defines a "scenario analysis" as an analysis of the impact of a combination of adverse events. The Monte Carlo simulation algorithm developed in this paper allows for a mathematically rigorous description how such scenarios can be generated, being flexible enough to cover also extreme situations.

In what follows, we shall focus mainly on the Non-Life Modules under Solvency II. Therefore we only consider continuous risk distributions. In this case, VaR is simply a quantile of the cumulative risk distribution function.

For corresponding considerations for the Life and Capital Asset Modules under Solvency II, we refer to Boonen (2017) or Varnell (2011).

Besides Solvency II aspects, the method proposed in this paper might also be of interest for reinsurance companies for the risk assessment of statistically dependent natural perils like windstorm, hail or flooding triggered by adverse climate conditions.

## 2. Unfavourable patchwork copulas

Patchwork copulas in the context of risk management have been treated in detail by Arbenz et al. (2012), Cottin and Pfeifer (2014), Pfeifer (2013), Pfeifer et al. (2016, 2017, 2019) and Hummel (2018), among others. In several of the cited papers the question of an unfavourable, i.e. superadditive VaR estimate for a portfolio of aggregated risks was in particular emphasized, see also Pfeifer and Ragulina (2018). However, the construction of worst VaR scenarios in this context is quite complicated; a numerical approach to a constructive solution is e.g. given by the rearrangement algorithm (see e.g. Arbenz et al. (2012), Embrechts et al. (2013) or Mainik (2015)). From a practical point of view, simpler and yet explicit constructions for unfavourable VaR estimates by appropriate copula constructions seem to be a useful alternative. In this paper, we describe how such a construction could be performed. We start with an explicit approach in two dimensions that is later extended to arbitrary dimensions. For better readability, all proofs are shifted to the appendix.

**Lemma 1.** Let, for $d \geq 2$, $d \in \mathbb{N}$, $\mathbf{U} = (U_1, \cdots, U_d)$ and $\mathbf{V} = (V_1, \cdots, V_d)$ be $d$-dimensional random vectors over $[0,1]^d$ with continuous uniform margins (i.e., $\mathbf{U}$ and $\mathbf{V}$ represent $d$-dimensional copulas). Let further $I$ denote a binomially distributed random variable, independent of $\mathbf{U}$ and $\mathbf{V}$, with $P(I=1) = p \in (0,1)$. Then the random vector $\mathbf{W}$ with components $W_i := I \cdot p \cdot U_i + (1-I) \cdot [p + (1-p) \cdot V_i]$ for $1 \leq i \leq d$ also has continuous uniform margins, i.e. $\mathbf{W}$ represents a $d$-dimensional copula.

Note that $\mathbf{W}$ can be considered as a special case of ordinal sums, cf. Nelsen (2006), chapter 3.3.2 for the two-dimensional case, and for the multivariate case, Jaworski and Rychlik (2008), relation (4.31), Jaworski (2009), Definition 2.1 and Durante and Sempi (2016), Example 2.2.10 and Chapter 3.8.

Suppose now that a portfolio of $d$ insurance risks is considered where a mutual probabilistic dependence structure is assumed to be described by $\mathbf{U}$. If the $d$ (for simplicity assumed continuous) marginal risk distribution functions are denoted by $F_1, \cdots, F_d$ and by $Q_1, \cdots, Q_d$ their pseudo-inverses (quantile functions), then both random vectors $(Q_1(U_1), \cdots, Q_d(U_d))$ and $(Q_1(W_1), \cdots, Q_d(W_d))$ represent a risk vector $\mathbf{X} = (X_1, \cdots, X_d)$ with the given marginal distributions. However, w.r.t. to risk aggregation, $\mathbf{X} := (Q_1(W_1), \cdots, Q_d(W_d))$ creates in general an unfavourable VaR scenario for $S = \sum_{i=1}^{d} X_i$, even if $p$ is close to 1 and therefore $\mathbf{U}$ and $\mathbf{W}$ differ only marginally. The following graph shows the corresponding support of $\mathbf{W}$ in two dimensions.

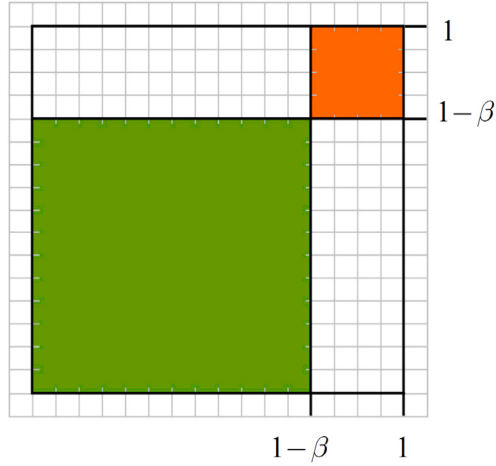

**Figure 2.** Shape of the support of the underlying patchwork copula

In the sequel put $p := 1-\beta$ for $0 < \beta < 1$. Then $\mathbf{W} = I \cdot (1-\beta) \cdot \mathbf{U} + (1-I) \cdot (1-\beta+\beta \cdot \mathbf{V})$.

We start with some further preliminary Lemmata.

**Lemma 2.** Let $\mathbf{W}_1 := (1-\beta) \cdot \mathbf{U}$, $\mathbf{W}_2 := 1-\beta+\beta \cdot \mathbf{V}$, $Z_{1i} := Q_i(W_{1i})$ and $Z_{2i} := Q_i(W_{2i})$, $i = 1, 2$. Then there hold

$$F_{Z_{1i}}(x,\beta) = \begin{cases} \dfrac{F_i(x)}{1-\beta}, & 0 \leq x \leq Q_i(1-\beta) \\ 1, & x \geq Q_i(1-\beta) \end{cases} \quad \text{and} \quad F_{Z_{2i}}(x,\beta) = \begin{cases} 0, & 0 \leq x \leq Q_i(1-\beta) \\ \dfrac{F_i(x)+\beta-1}{\beta}, & x \geq Q_i(1-\beta). \end{cases}$$

**Lemma 3.** Assume that $f$ and $g$ are Lebesgue densities of independent random variables $X$ and $Y$, concentrated on the same finite interval $[0, M]$ with $M > 0$. Then $S := X + Y$ has the density $h_1$ given by

$$h_1(x) = \begin{cases} \displaystyle\int_0^x f(x-y)g(y)\,dy, & 0 \leq x \leq M \\ \displaystyle\int_{x-M}^M f(x-y)g(y)\,dy, & M \leq x \leq 2M. \end{cases}$$

If $f$ and $g$ are concentrated on the same infinite interval $[M, \infty)$ with $M \geq 0$, then $S := X + Y$ has the density $h_2$ given by

$$h_2(x) = \int_M^{x-M} f(x-y)g(y)\,dy, \quad x \geq 2M.$$

In particular, if $F$ and $G$ are the corresponding cdf's pertaining to $f$ and $g$, resp., then in either case, $\left.\frac{d}{dx} F * G(x)\right|_{x=2M} = 0$, where $*$ means convolution.

**Lemma 4.** Assume that all $F_i \equiv F$ being equal with quantile function $Q$, and that $\mathbf{U}$ and $\mathbf{V}$ have independent components each. Denote

$$\underline{F}(x,\beta) := \begin{cases} \frac{F(x)}{1-\beta}, & x \leq Q(1-\beta) \\ 1, & x \geq Q(1-\beta) \end{cases} \quad \text{and} \quad \overline{F}(x,\beta) := \frac{F(x+Q(1-\beta))+\beta-1}{\beta}, \; x \geq 0.$$

Let further denote $X_i := Q(W_i)$ and $S := \sum_{i=1}^{d} X_i$. Then we can conclude that

$$F_S(x,\beta) = \begin{cases} (1-\beta)\underline{F}^{d*}(x,\beta), & x \leq dQ(1-\beta) \\ (1-\beta)+\beta\overline{F}^{d*}(x-dQ(1-\beta),\beta), & x > dQ(1-\beta), \end{cases}$$

where $*$ again means convolution. If $F$ has a density $f$, then correspondingly

$$\underline{f}(x,\beta) := \begin{cases} \frac{f(x)}{1-\beta}, & x \leq Q(1-\beta) \\ 0, & x \geq Q(1-\beta), \end{cases} \quad \overline{f}(x,\beta) := \frac{f(x+Q(1-\beta))}{\beta}, \; x \geq 0$$

and

$$f_S(x,\beta) = \begin{cases} (1-\beta)\underline{f}^{d*}(x,\beta), & x \leq dQ(1-\beta) \\ (1-\beta)+\beta\overline{f}^{d*}(x-dQ(1-\beta),\beta), & x > dQ(1-\beta). \end{cases}$$

The following examples show the effect of a risk aggregation with an unfavourable VaR scenario for two dimensions in detail.

**Example 1** (exponential distributions). Assume that $F_1 = F_2 = \begin{cases} 0, & x < 0 \\ 1-e^{-x}, & x \geq 0. \end{cases}$ Then

$$F_{Z_{1i}}(x,\beta) = \frac{1-e^{-x}}{1-\beta}, 0 \leq x \leq -\ln(\beta) \text{ and } F_{Z_{2i}}(x,\beta) = \frac{\beta-e^{-x}}{\beta} = 1-e^{-x-\ln(\beta)}, x \geq -\ln(\beta), i=1,2.$$

For the corresponding densities, we obtain by differentiation

$$f_{Z_{1i}}(x,\beta) = \begin{cases} \frac{e^{-x}}{1-\beta}, & 0 \leq x \leq -\ln(\beta) \\ 0, & x > -\ln(\beta), \end{cases} \quad f_{Z_{2i}}(x,\beta) = \begin{cases} 0, & x < -\ln(\beta) \\ e^{-x-\ln(\beta)}, & x \geq -\ln(\beta) \end{cases}, i=1,2$$

and

$$\underline{f}(x,\beta) = \begin{cases} \dfrac{e^{-x}}{1-\beta}, & 0 \leq x \leq -\ln(\beta) \\ 0, & x > -\ln(\beta), \end{cases} \qquad \overline{f}(x,\beta) = \begin{cases} 0, & x < 0 \\ e^{-x}, & x \geq 0. \end{cases}$$

By Lemma 4, we obtain the following density $f_S$ of the aggregated risk $S$:

$$f_S(x,\beta) = \begin{cases} \dfrac{xe^{-x}}{1-\beta}, & 0 \leq x \leq -\ln(\beta) \\ \dfrac{(-2\ln(\beta)-x)e^{-x}}{1-\beta}, & -\ln(\beta) \leq x \leq -2\ln(\beta) \\ \dfrac{(x+2\ln(\beta))e^{-x}}{\beta} & x \geq -2\ln(\beta) \end{cases}$$

with the corresponding cdf $F_S$:

$$F_S(x,\beta) = \begin{cases} \dfrac{1-(1+x)e^{-x}}{1-\beta}, & 0 \leq x \leq -\ln(\beta) \\ \dfrac{1-2\beta+2e^{-x}\ln(\beta)+(1+x)e^{-x}}{1-\beta}, & -\ln(\beta) \leq x \leq -2\ln(\beta) \\ \dfrac{\beta - 2e^{-x}\ln(\beta)-(1+x)e^{-x}}{\beta} & x \geq -2\ln(\beta). \end{cases}$$

For the following graph, let $g$ denote the density of $T := Q_1(U_1) + Q_2(U_2)$ (independent summands, Gamma distribution).

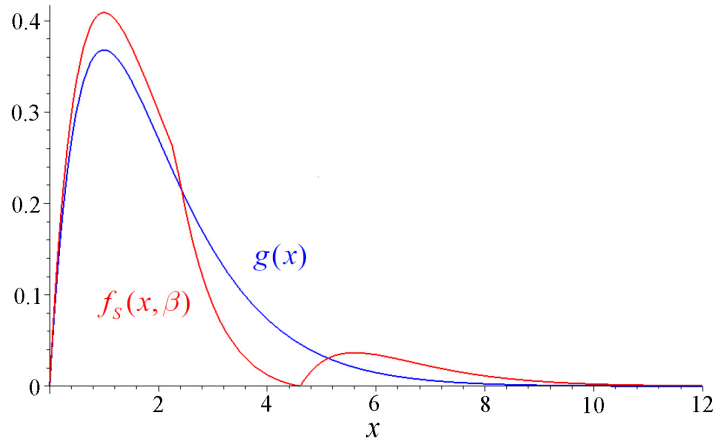

**Figure 3.** Plot of the densities $f_S(x,\beta)$ for $\beta = 0.1$ and $g(x)$, Example 1

In what follows, let $G$ denote the cdf of $T := Q_1(U_1) + Q_2(U_2)$ (independent summands, Gamma distribution) and $H$ the cdf of $S$ under the worst VaR scenario, i.e. the distribution of **V** corresponds to the lower Fréchet bound or countermonotonicity copula (see e.g. Embrechts et al. (2013), Remark 3 and the comments after Fig. 3, or Pfeifer (2013)). In this case we have

$$H(x,\beta) = \begin{cases} F_S(x), & x \leq -2\ln(\beta) \\ 1-\beta, & -2\ln(\beta) \leq x \leq -2\ln(\beta/2) \\ 1-\beta+\sqrt{\beta^2-4e^{-x}}, & x \geq -2\ln(\beta/2). \end{cases}$$

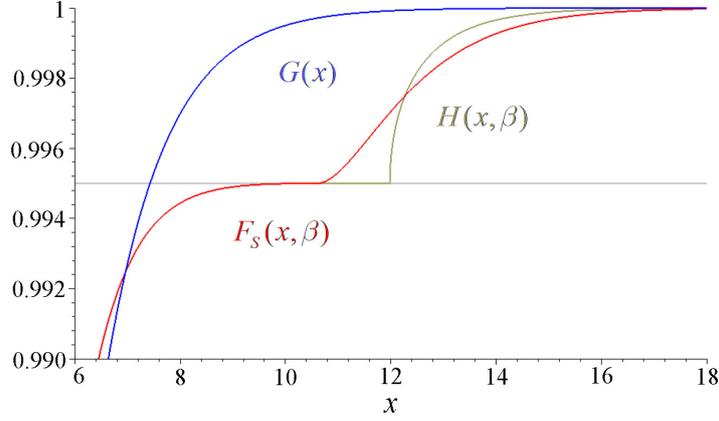

**Figure 4.** Plots of the cdf's $F_S(x,\beta)$, $G(x)$ and $H(x,\beta)$, for $\beta = 0.005$, Example 1

Note that with the Solvency II standard $\alpha = 0.005$, we get here, for $\beta = \alpha$, $\text{VaR}_\alpha(S) = 10.5914 > \text{VaR}_\alpha(T) = 7.4301$. For the worst VaR scenario, however, we get $\text{wVaR}_\alpha(S) = 11.9829 > 10.5966 = \text{SVaR}_\alpha := \text{VaR}_\alpha(X_1) + \text{VaR}_\alpha(X_2) > \text{VaR}_\alpha(S) = 10.5914$. This means that even with the construction for $S$ with $\beta = \alpha$, we still have a (quite small) diversification effect, but not in the worst VaR scenario. This changes, however, if we look at $\text{VaR}_\alpha(S) = 10.9630$ when we replace $\beta$ by $\alpha + \varepsilon$ in the definition of **W** for e.g. $\varepsilon = 0.001$.

The following graph shows the cdf's in the tails for several choices of $\varepsilon$.

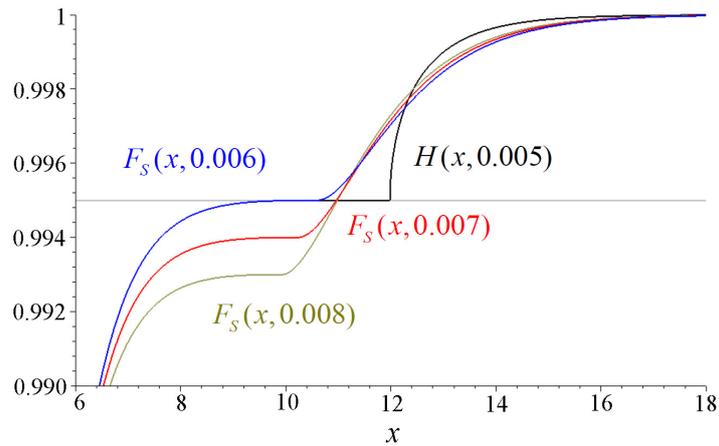

**Figure 5.** Plots of the cdf's $F_S(x, 0.005+\varepsilon)$ for $\varepsilon \in \{0.001, 0.002, 0.003\}$ and $H(x, 0.005)$, Example 1

The following graph shows the values of $Q_S(0.995, \beta) = F_S^{-1}(0.995, \beta)$ in the range $0.0062 \leq \beta \leq 0.0076$.

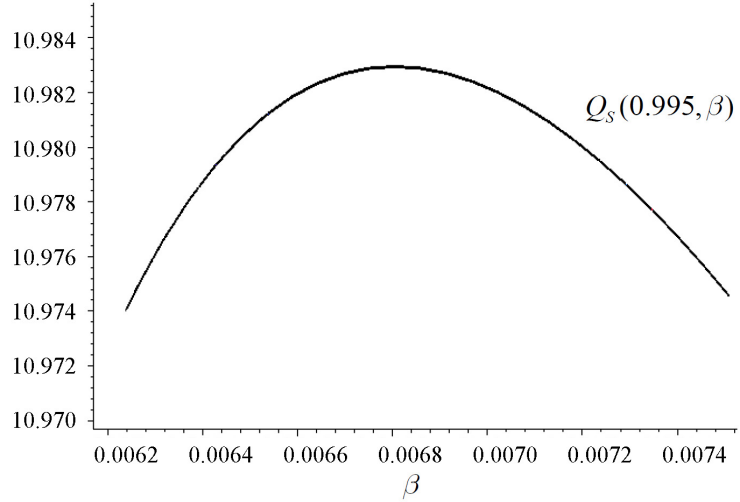

**Figure 6.** Plot of the parametrized quantile function $Q_S(0.995, \beta) = F_S^{-1}(0.995, \beta)$, Example 1

A numerical calculation shows that for $\alpha = 0.005$ the maximal $\text{VaR}_\alpha(S) = 10.9829$ is attained for $\beta = 0.0068$, i.e. $\varepsilon = 0.0018$.

The following table summarizes the results found, for $\alpha = 0.005$.

**Table 1.** Summarized results for Example 1

|  | $\beta$ | | | | |
|---|---|---|---|---|---|
|  | 0.0050 | 0.0060 | 0.0068 | 0.0070 | 0.0080 |
| $\text{VaR}_\alpha(S)$ | 10.5914 | 10.9630 | 10.9829 | 10.9821 | 10.9618 |
| $\text{VaR}_\alpha(T)$ | 7.4301 | 7.4301 | 7.4301 | 7.4301 | 7.4301 |
| $\text{wVaR}_\alpha(S)$ | 11.9829 | 11.9829 | 11.9829 | 11.9829 | 11.9829 |
| $\text{SVaR}_\alpha$ | 10.5966 | 10.5966 | 10.5966 | 10.5966 | 10.5966 |

**Example 2** (uniform distributions). Assume that $F_1 = F_2 = \begin{cases} 0, & x \leq 0 \\ x, & 0 \leq x \leq 1 \\ 1, & x \geq 1. \end{cases}$ Then

$F_{Z_{1i}}(x, \beta) = \dfrac{x}{1-\beta}$, $0 \leq x \leq 1-\beta$ and $F_{Z_{2i}}(x, \beta) = \dfrac{x+\beta-1}{\beta}$, $x \geq 1-\beta$, $i = 1, 2$.

By Lemma 4, we obtain the following density $f_S$ of the aggregated risk $S$:

$$f_S(x,\beta) = \begin{cases} \dfrac{x}{1-\beta}, & x \leq 1-\beta \\[4pt] \dfrac{2-2\beta-x}{1-\beta}, & 1-\beta \leq x \leq 2-2\beta \\[4pt] \dfrac{x-2+2\beta}{\beta}, & 2-2\beta \leq x \leq 2-\beta \\[4pt] \dfrac{2-x}{\beta}, & 2-\beta \leq x \leq 2 \end{cases}$$

with the corresponding cdf $F_S$ :

$$F_S(x,\beta) = \begin{cases} \dfrac{x^2}{2(1-\beta)}, & x \leq 1-\beta \\[4pt] \dfrac{4x(1-\beta)-x^2-2(1-\beta)^2}{2(1-\beta)}, & 1-\beta \leq x \leq 2-2\beta \\[4pt] \dfrac{4(1-\beta)(1-x)+x^2-2\beta+2\beta^2}{2\beta}, & 2-2\beta \leq x \leq 2-\beta \\[4pt] \dfrac{2\beta-4(1-x)-x^2}{2\beta}, & 2-\beta \leq x \leq 2. \end{cases}$$

In what follows $g$ is the density of $T := Q_1(U_1) + Q_2(U_2)$ (independent summands, triangle distribution).

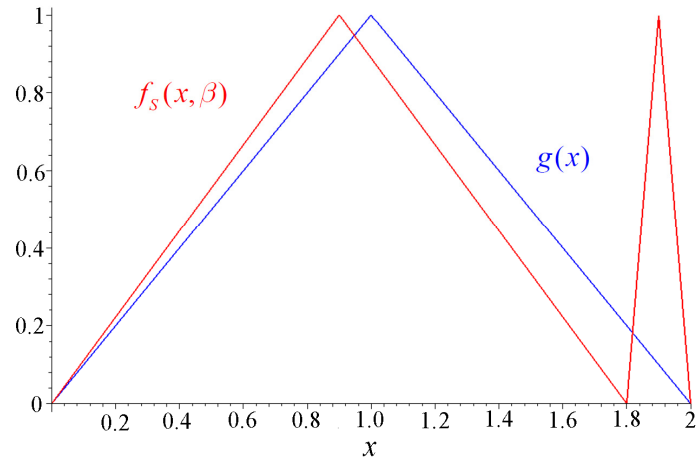

**Figure 7.** Plot of the densities $f_S(x,\beta)$ for $\beta = 0.1$ and $g(x)$, Example 2

In the following graph $G$ is the cdf for $T := Q_1(U_1) + Q_2(U_2)$ (independent summands, triangle distribution) and $H$ the cdf for $S$ under the worst VaR scenario, i.e. the distribution of **V** corresponds to the lower Fréchet bound. In this case we have

$$H(x,\beta) = \begin{cases} F_S(x) & x \leq 2-2\beta \\ 1-\beta, & 2-2\beta \leq x < 2-\beta \\ 1, & x \geq 2-\beta. \end{cases}$$

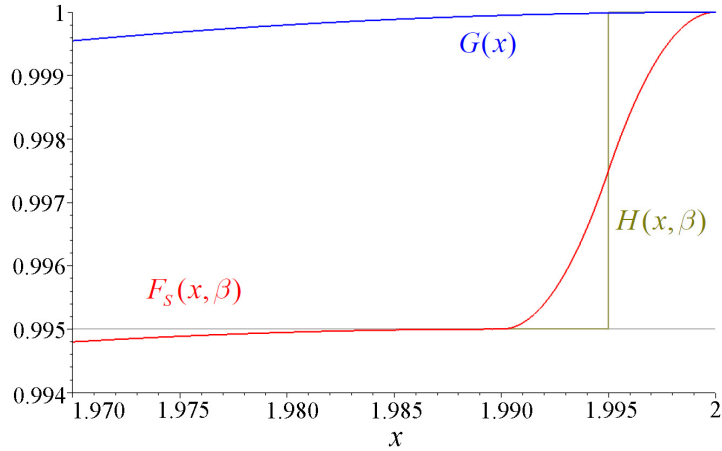

**Figure 8.** Plots of the cdf's $F_S(x,\beta)$, $G(x)$, and $H(x,\beta)$ for $\beta = 0.005$, Example 2

Note that with the Solvency II standard $\alpha = 0.005$, we have here, for $\beta = \alpha$, $\text{VaR}_\alpha(S) = 1.9900 = \text{VaR}_\alpha(T)$. For the worst VaR scenario, however, we get here $\text{wVaR}_\alpha(S) = 1.9950 > 1.9900 = \text{VaR}_\alpha(X_1) + \text{VaR}_\alpha(X_2) = \text{SVaR}_\alpha = \text{VaR}_\alpha(S)$. This means that with the construction for $S$ we have no true diversification effect, in contrast to the worst VaR scenario. This changes, however, if we look at $\text{VaR}_\alpha(S) = 1.9910$ when we replace $\beta$ by $\alpha + \varepsilon$ in the definition of **W** for e.g. $\varepsilon = 0.001$.

The following graph shows the cdf's in the tails for several choices of $\varepsilon$.

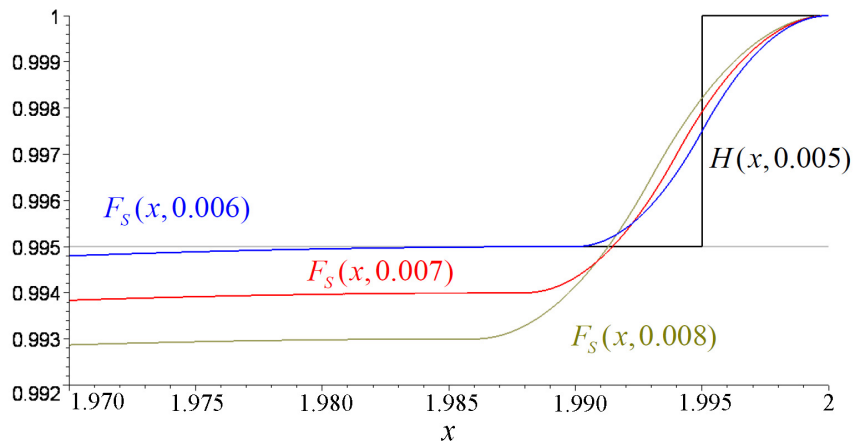

**Figure 9.** Plots of the cdf's $F_S(x, 0.005+\varepsilon)$ for $\varepsilon \in \{0.001, 0.002, 0.003\}$ and $H(x, 0.005)$, Example 2

The following graph shows the values of $Q_S(0.995, \beta) = F_S^{-1}(0.995, \beta)$ in the range $0.0054 \leq \beta \leq 0.007$.

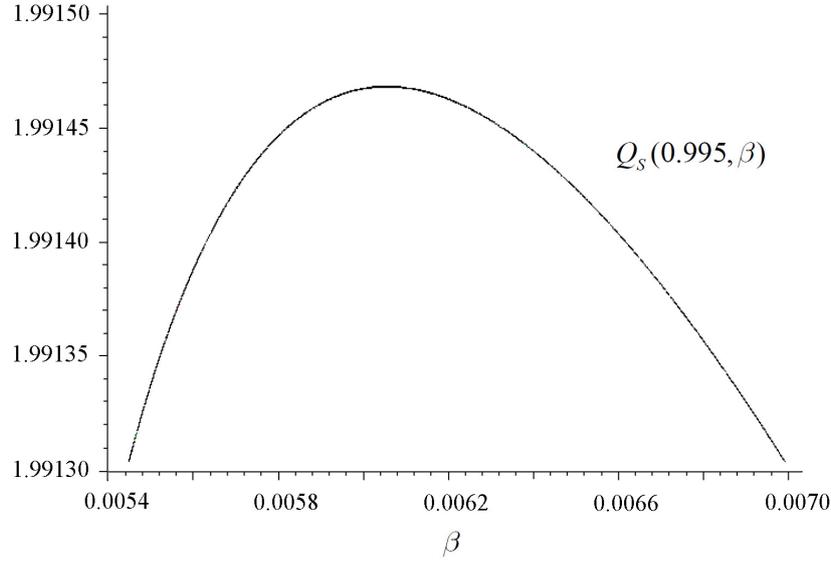

**Figure 10.** Plot of the parametrized quantile function $Q_S(0.995, \beta) = F_S^{-1}(0.995, \beta)$, Example 2

A numerical calculation shows that for $\alpha = 0.005$ the maximal $\text{VaR}_\alpha(S) = 1.9915$ is attained for $\beta = 0.0060$, i.e. $\varepsilon = 0.0010$.

Note that in this example a closed-form representation for $Q_S(u, \beta)$ is given by

$Q_S(u, \beta) = 2 - 2\beta + \sqrt{2\beta(\beta + u - 1)}, 1 - \beta \leq u \leq 1 - \dfrac{\beta}{2}$. This implies

$Q_S(1 - \alpha, \beta) = 2 - 2\beta + \sqrt{2\beta(\beta - \alpha)}, \alpha \leq \beta \leq 2\alpha$

with its maximum being attained for $\beta_0 = \dfrac{1 + \sqrt{2}}{2} \alpha$ with value

$Q_S(1 - \alpha, \beta_0) = 2 - \left(1 + \dfrac{\sqrt{2}}{2}\right)\alpha.$

Note that in contrast the worst VaR here is $\text{wVaR}_\alpha(S) = 2 - \alpha$.

The following table summarizes the results found, for $\alpha = 0.005$.

**Table 2.** Summarized results for Example 2

| | $\beta$ | | | | |
|---|---|---|---|---|---|
| | 0.0050 | 0.0055 | 0.0060 | 0.0065 | 0.0070 |
| $\text{VaR}_\alpha(S)$ | 1.9900 | 1.9130 | 1.9915 | 1.9914 | 1.9913 |
| $\text{VaR}_\alpha(T)$ | 1.9900 | 1.9900 | 1.9900 | 1.9900 | 1.9900 |
| $\text{wVaR}_\alpha(S)$ | 1.9950 | 1.9950 | 1.9950 | 1.9950 | 1.9950 |
| $\text{SVaR}_\alpha$ | 1.9900 | 1.9900 | 1.9900 | 1.9900 | 1.9900 |

**Example 3** (Pareto distributions). Assume that $F_1 = F_2 = \begin{cases} 0, & x \leq 0 \\ \dfrac{x}{1+x}, & x > 0. \end{cases}$ Then

$$F_{Z_{1i}}(x,\beta) = \frac{x}{(1-\beta)(1+x)}, \quad 0 \leq x \leq \frac{1}{\beta} - 1 \text{ and } F_{Z_{2i}}(x,\beta) = 1 - \frac{1}{\beta(1+x)}, \quad x \geq \frac{1}{\beta} - 1, \quad i = 1, 2.$$

For the corresponding densities, we obtain by differentiation

$$f_{Z_{1i}}(x,\beta) = \begin{cases} \dfrac{1}{(1-\beta)(1+x)^2}, & 0 \leq x \leq \dfrac{1}{\beta} - 1 \\ 0, & x > \dfrac{1}{\beta} - 1, \end{cases} \qquad f_{Z_{2i}}(x,\beta) = \begin{cases} 0, & x < \dfrac{1}{\beta} - 1 \\ \dfrac{1}{\beta(1+x)^2}, & x \geq \dfrac{1}{\beta} - 1 \end{cases}$$

and

$$\underline{f}(x,\beta) = \begin{cases} \dfrac{1}{(1-\beta)(1+x)^2}, & 0 \leq x \leq \dfrac{1}{\beta} - 1 \\ 0, & x > \dfrac{1}{\beta} - 1, \end{cases} \qquad \overline{f}(x,\beta) = \begin{cases} 0, & x < 0 \\ \dfrac{\beta}{(1+\beta x)^2}, & x \geq 0. \end{cases}$$

In order to calculate the density $f_S$ of the aggregated risk $S$, we need a suitable partial fraction representation of $\underline{f}(x-y)\underline{f}(y)$ and $\overline{f}(x-y)\overline{f}(y)$. Note that in general, we have

$$\frac{1}{(1+x-y)(1+y)} = \frac{1}{2+x} \left[ \frac{1}{1+x-y} + \frac{1}{1+y} \right]$$

and

$$\frac{1}{(1+x-y)^2(1+y)^2} = \frac{1}{(2+x)^2}\left[\frac{1}{(1+x-y)} + \frac{1}{(1+y)}\right]^2$$

$$= \frac{1}{(2+x)^2}\left[\frac{1}{(1+x-y)^2} + \frac{1}{(1+y)^2} + \frac{2}{2+x}\left[\frac{1}{1+x-y} + \frac{1}{1+y}\right]\right]$$

from which we obtain, by Lemma 4,

$$F_S(x,\beta) = \begin{cases} \dfrac{x^2 + 2x - 2\ln(1+x)}{(2+x)^2(1-\beta)}, & 0 \leq x \leq \dfrac{1}{\beta} - 1 \\ \dfrac{(1-2\beta)x^2 + (4-6\beta)x - 4\beta + 4 + 2\ln(\beta x + 2\beta - 1)}{(2+x)^2(1-\beta)}, & \dfrac{1}{\beta} - 1 \leq x \leq 2\left(\dfrac{1}{\beta} - 1\right) \\ \dfrac{x^2 - 2x + \dfrac{2}{\beta}\ln(\beta x + 2\beta - 1)}{(2+x)^2} & x \geq 2\left(\dfrac{1}{\beta} - 1\right). \end{cases}$$

The density $f_S(x)$ follows by differentiation.

In the following $g$ denotes the density of $T := Q_1(U_1) + Q_2(U_2)$ (independent summands).

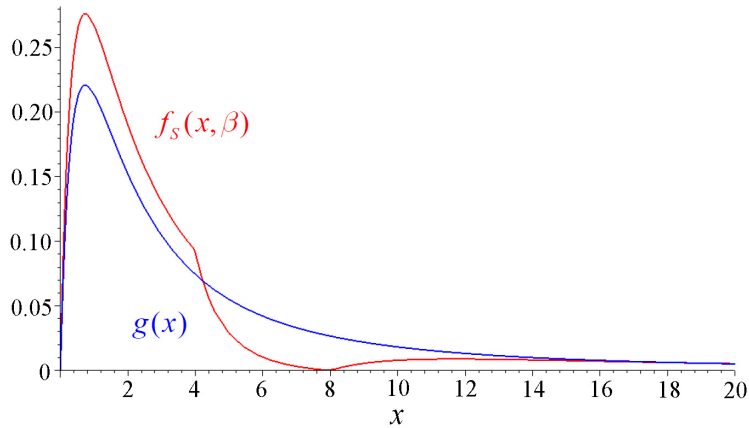

**Figure 11.** Plots of densities $f_S(x,\beta)$ for $\beta = 0.1$ and $g(x)$, Example 3

In the following graph $G$ is the cdf of $T := Q_1(U_1) + Q_2(U_2)$ (independent summands) and $H$ the cdf of $S$ under the worst VaR scenario, i.e. the distribution of **V** corresponds again to the lower Fréchet bound. In this case we have

$$H(x,\beta) = \begin{cases} F_S(x,\beta), & x \leq \dfrac{2}{\beta} - 2 \\ 1-\beta, & \dfrac{2}{\beta} - 2 \leq x \leq \dfrac{4}{\beta} - 2 \\ 1-\beta + \sqrt{\beta^2 - \dfrac{4\beta}{2+x}}, & x \geq \dfrac{4}{\beta} - 2. \end{cases}$$

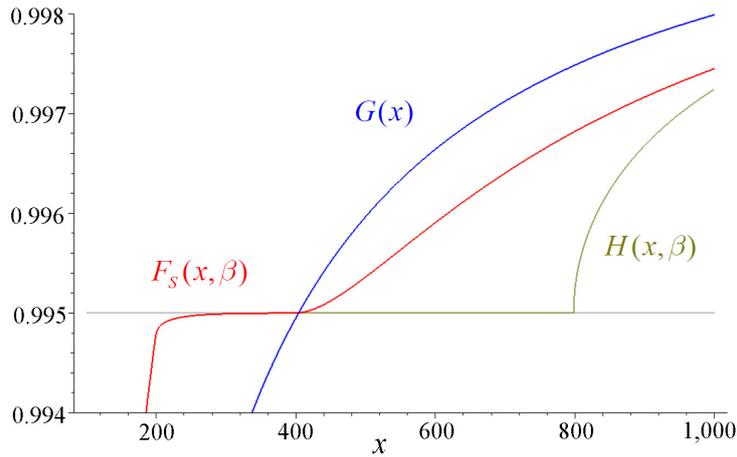

**Figure 12.** Plots of the cdf's $F_S(x,\beta)$, $G(x)$ and $H(x,\beta)$ for $\beta = 0.005$, Example 3

Note that with the Solvency II standard $\alpha = 0.005$, we have here, for $\beta = \alpha$, $\text{VaR}_\alpha(S) = 397.3168 < \text{VaR}_\alpha(T) = 403.9161$. For the worst VaR scenario, however, we get $\text{wVaR}_\alpha(S) = 798 > 398 = \text{VaR}_\alpha(X_1) + \text{VaR}_\alpha(X_2) = \text{SVaR}_\alpha > \text{VaR}_\alpha(S) = 397.3168$. This means that even with the construction for $S$ we still have a (quite small) diversification effect, but not in the worst VaR scenario, as expected. This changes, however, if we look at $\text{VaR}_\alpha(S) = 488.2116$ when we replace $\beta$ by $\beta + \varepsilon$ in the definition of **W** for e.g. $\varepsilon = 0.001$.

The following graph shows the cdf's in the tail for several choices of $\varepsilon$.

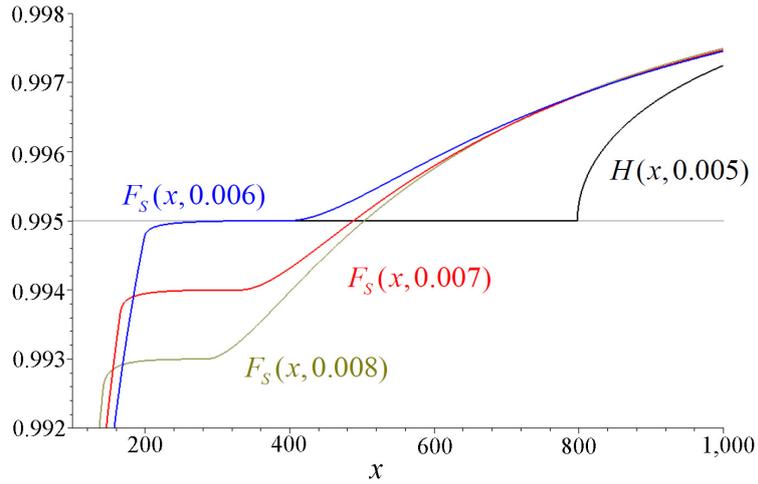

**Figure 13.** Plots of the cdf's $F_S(x, 0.005 + \varepsilon)$ for $\varepsilon \in \{0.001, 0.002, 0.003\}$ and $H(x, 0.005)$, Example 3

The following graph shows the values of $Q_S(0.995, \beta) = F_S^{-1}(0.995, \beta)$ in the range $0.007 \leq \beta \leq 0.012$.

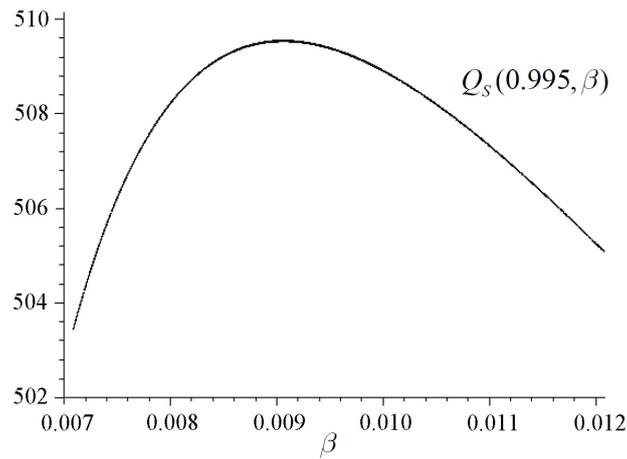

**Figure 14.** Plot of the parametrized quantile function $Q_S(0.995, \beta) = F_S^{-1}(0.995, \beta)$, Example 3

A numerical calculation shows that for $\alpha = 0.005$ the maximum $\text{VaR}_\alpha(S) = 509.3804$ is attained for $\beta = 0.0089$, i.e. $\varepsilon = 0.0039$.

Table 3. Summarized results for Example 3

| | $\beta$ | | | | |
|---|---|---|---|---|---|
| | 0.0050 | 0.0070 | 0.0089 | 0.0100 | 0.0110 |
| $\text{VaR}_\alpha(S)$ | 397.3168 | 503.2848 | 509.3804 | 508.6489 | 507.0076 |
| $\text{VaR}_\alpha(T)$ | 403.9161 | 403.9161 | 403.9161 | 403.9161 | 403.9161 |
| $\text{wVaR}_\alpha(S)$ | 798.0000 | 798.0000 | 798.0000 | 798.0000 | 798.0000 |
| $\text{SVaR}_\alpha$ | 398.0000 | 398.0000 | 398.0000 | 398.0000 | 398.0000 |

These examples show that it is generally possible to obtain unfavourable VaR scenarios by a suitable choice of $\beta = \alpha + \varepsilon$ in the definition of $\mathbf{W}$, i.e. scenarios which lead to an opposite diversification effect in the portfolio and which are sometimes close to the worst VaR scenario.

We continue with a particular construction of $\mathbf{W}$ which allows in general for an unfavourable VaR scenario.

**Lemma 5.** For $d \in \mathbb{N}$, $d > 1$ let $\mathbf{I}_d$ denote the $d$-dimensional unit matrix, $\mathbf{e}_d = (1, \cdots, 1)$ the $d$-dimensional row vector consisting of just ones, and $\mathbf{E}_d = \mathbf{e}_d^{tr} \mathbf{e}_d$ the $d \times d$ matrix with all entries equal to unity. Then $\Sigma_d = (1-r)\mathbf{I}_d + r\mathbf{E}_d$ is a correlation matrix iff $-\frac{1}{d-1} \leq r \leq 1$. In the general case, the latent roots $\lambda_i$ of $\Sigma_d$ are given by $\lambda_1 = 1 + (d-1)r$ and $\lambda_i = 1 - r$, $i = 2, \cdots, d$. An orthonormal basis $T_1, \cdots, T_d$ of corresponding latent vectors is given by $T_1 = \frac{1}{\sqrt{d}} \mathbf{e}_d^{tr}$ and $T_j = (t_{1j}, \cdots, t_{dj})^{tr}$ for $2 \leq j \leq d$ where

$$t_{ij} = \begin{cases} -\frac{1}{\sqrt{j(j-1)}}, & 1 \leq i < j \\ \sqrt{\frac{j-1}{j}}, & j = i \\ 0, & i > j. \end{cases}$$

Hence $\Sigma_d$ possesses the spectral decomposition $\Sigma_d = \mathbf{A}\mathbf{A}^{tr}$ with $\mathbf{A} = \mathbf{T}\sqrt{\Delta}$ where $\mathbf{T} = [T_1, \cdots, T_d]$ and $\Delta = \text{diag}(\lambda_1, \cdots, \lambda_d)$.

Note that there is also an alternative possibility to represent latent roots $\lambda_j^*$ and normalized latent vectors $T_j^* = (t_{1j}^*, \cdots, t_{dj}^*)^{tr}$, $j = 1, \cdots, d$ of $\Sigma_d$ since $\Sigma_d$ is a particular symmetric

Toeplitz matrix for which the latent roots and normalized latent vectors can be expressed via trigonometric functions, see e.g. Basilevsky (1983), relations (5.89) and (5.90). In particular, we can choose

$$\lambda_j^* = 1 + r \sum_{i=1}^{d-1} \cos\left(\frac{2\pi ij}{d}\right) = \begin{cases} 1-r, & j=1,\cdots,d-1 \\ 1+(d-1)r, & j=d \end{cases}$$

and

$$t_{ij}^* = \frac{\cos\left(\frac{2\pi ij}{d}\right) + \sin\left(\frac{2\pi ij}{d}\right)}{\sqrt{d}}, \ 1 \leq i,j \leq d.$$

This is due to the fact that the latent roots have multiplicities, hence the linear space spanned by the corresponding latent vectors is $d-1$-dimensional, allowing for different representations of the corresponding linear basis. However, for our purposes, the representation in Lemma 5 is more suited.

In what follows we will call a Gaussian copula derived from the correlation matrix $\Sigma_d = \frac{d}{d-1}\mathbf{I}_d - \frac{1}{d-1}\mathbf{E}_d$ for $r = -\frac{1}{d-1}$ a *minimal correlation Gaussian copula*. Note that the corresponding multivariate normal distribution is degenerated since $\Sigma_d$ is singular, i.e. a random vector $\mathbf{X}$ with zero mean and correlation matrix $\Sigma_d$ has the representation $\mathbf{X} = A\mathbf{Y}$ where $\mathbf{Y}$ has a standard multivariate normal distribution with mean zero and variance-covariance matrix $\mathbf{I}_d$. For $d = 2$, the minimal correlation Gaussian copula is identical to the lower Fréchet bound or countermonotonicity copula.

## 3. A case study

The following example shows the effects of such an approach for the 19-dimensional data set discussed in Pfeifer et al. (2019). It contains insurance losses from a non-life portfolio of natural perils in $d = 19$ areas in central Europe over a time period of 20 years. The losses are given in million monetary units (MMU).

**Table 4.** Insurance losses from a Nat-Cat portfolio in central Europe

| Year | Area 1 | Area 2 | Area 3 | Area 4 | Area 5 | Area 6 | Area 7 | Area 8 | Area 9 | Area 10 |
|---|---|---|---|---|---|---|---|---|---|---|
| 1 | 23.664 | 154.664 | 40.569 | 14.504 | 10.468 | 7.464 | 22.202 | 17.682 | 12.395 | 18.551 |
| 2 | 1.080 | 59.545 | 3.297 | 1.344 | 1.859 | 0.477 | 6.107 | 7.196 | 1.436 | 3.720 |
| 3 | 21.731 | 31.049 | 55.973 | 5.816 | 14.869 | 20.771 | 3.580 | 14.509 | 17.175 | 87.307 |
| 4 | 28.99 | 31.052 | 30.328 | 4.709 | 0.717 | 3.530 | 6.032 | 6.512 | 0.682 | 3.115 |
| 5 | 53.616 | 62.027 | 57.639 | 1.804 | 2.073 | 4.361 | 46.018 | 22.612 | 1.581 | 11.179 |
| 6 | 29.95 | 41.722 | 12.964 | 1.127 | 1.063 | 4.873 | 6.571 | 11.966 | 15.676 | 24.263 |
| 7 | 3.474 | 14.429 | 10.869 | 0.945 | 2.198 | 1.484 | 4.547 | 2.556 | 0.456 | 1.137 |
| 8 | 10.02 | 31.283 | 21.116 | 1.663 | 2.153 | 0.932 | 25.163 | 3.222 | 1.581 | 5.477 |
| 9 | 5.816 | 14.804 | 128.072 | 0.523 | 0.324 | 0.477 | 3.049 | 7.791 | 4.079 | 7.002 |
| 10 | 170.725 | 576.767 | 108.361 | 41.599 | 20.253 | 35.412 | 126.698 | 71.079 | 21.762 | 64.582 |
| 11 | 21.423 | 50.595 | 4.360 | 0.327 | 1.566 | 64.621 | 5.650 | 1.258 | 0.626 | 3.556 |
| 12 | 6.38 | 28.316 | 3.740 | 0.442 | 0.736 | 0.470 | 3.406 | 7.859 | 0.894 | 3.591 |
| 13 | 124.665 | 33.359 | 14.712 | 0.321 | 0.975 | 2.005 | 3.981 | 4.769 | 2.006 | 1.973 |
| 14 | 20.165 | 49.948 | 17.658 | 0.595 | 0.548 | 29.35 | 6.782 | 4.873 | 2.921 | 6.394 |
| 15 | 78.106 | 41.681 | 13.753 | 0.585 | 0.259 | 0.765 | 7.013 | 9.426 | 2.18 | 3.769 |
| 16 | 11.067 | 444.712 | 365.351 | 99.366 | 8.856 | 28.654 | 10.589 | 13.621 | 9.589 | 19.485 |
| 17 | 6.704 | 81.895 | 14.266 | 0.972 | 0.519 | 0.644 | 8.057 | 18.071 | 5.515 | 13.163 |
| 18 | 15.55 | 277.643 | 26.564 | 0.788 | 0.225 | 1.230 | 26.800 | 64.538 | 2.637 | 80.711 |
| 19 | 10.099 | 18.815 | 9.352 | 2.051 | 1.089 | 6.102 | 2.678 | 4.064 | 2.373 | 2.057 |
| 20 | 8.492 | 138.708 | 46.708 | 3.680 | 1.132 | 1.698 | 165.600 | 7.926 | 2.972 | 5.237 |

| Year | Area 11 | Area 12 | Area 13 | Area 14 | Area 15 | Area 16 | Area 17 | Area 18 | Area 19 |
|---|---|---|---|---|---|---|---|---|---|
| 1 | 1.842 | 4.100 | 46.135 | 14.698 | 44.441 | 7.981 | 35.833 | 10.689 | 7.299 |
| 2 | 0.429 | 1.026 | 7.469 | 7.058 | 4.512 | 0.762 | 14.474 | 9.337 | 0.740 |
| 3 | 0.209 | 2.344 | 22.651 | 4.117 | 26.586 | 3.920 | 13.804 | 2.683 | 3.026 |
| 4 | 0.521 | 0.696 | 31.126 | 1.878 | 29.423 | 6.394 | 18.064 | 1.201 | 0.894 |
| 5 | 2.715 | 1.327 | 40.156 | 4.655 | 104.691 | 28.579 | 17.832 | 1.618 | 3.402 |
| 6 | 4.832 | 0.701 | 16.712 | 11.852 | 29.234 | 7.098 | 17.866 | 5.206 | 5.664 |
| 7 | 0.268 | 0.580 | 11.851 | 2.057 | 11.605 | 0.282 | 16.925 | 2.082 | 1.008 |
| 8 | 0.741 | 0.369 | 3.814 | 1.869 | 8.126 | 1.032 | 14.985 | 1.390 | 1.703 |
| 9 | 0.524 | 6.554 | 5.459 | 3.007 | 8.528 | 1.920 | 5.638 | 2.149 | 2.908 |
| 10 | 9.882 | 6.401 | 106.197 | 44.912 | 191.809 | 90.559 | 154.492 | 36.626 | 36.276 |
| 11 | 1.052 | 8.277 | 22.564 | 8.961 | 19.817 | 16.437 | 25.990 | 2.364 | 6.434 |
| 12 | 0.136 | 0.364 | 28.000 | 7.574 | 3.213 | 1.749 | 12.735 | 1.744 | 0.558 |
| 13 | 1.990 | 15.176 | 57.235 | 23.686 | 110.035 | 17.373 | 7.276 | 2.494 | 0.525 |
| 14 | 0.630 | 0.762 | 25.897 | 3.439 | 8.161 | 3.327 | 24.733 | 2.807 | 1.618 |
| 15 | 0.770 | 15.024 | 36.068 | 1.613 | 6.127 | 8.103 | 12.596 | 4.894 | 0.822 |
| 16 | 0.287 | 0.464 | 24.211 | 38.616 | 51.889 | 1.316 | 173.080 | 3.557 | 11.627 |
| 17 | 0.590 | 2.745 | 16.124 | 2.398 | 20.997 | 2.515 | 5.161 | 2.840 | 3.002 |
| 18 | 0.245 | 0.217 | 12.416 | 4.972 | 59.417 | 3.762 | 24.603 | 7.404 | 19.107 |
| 19 | 0.415 | 0.351 | 10.707 | 2.468 | 10.673 | 1.743 | 27.266 | 1.368 | 0.644 |
| 20 | 0.566 | 0.708 | 22.646 | 6.652 | 14.437 | 63.692 | 113.231 | 7.218 | 2.548 |

A statistical analysis of the data shows a good fit to lognormal $\mathcal{LN}(\mu,\sigma)$-distributions for the losses per Area $k$, $k=1,\cdots,19$. The parameters $\mu_k$ and $\sigma_k$ for Area $k$ were hence estimated from the log data by calculating means and standard deviations.

**Table 5.** Distributional parameters for fitted lognormal loss distributions

| Parameter | Area 1 | Area 2 | Area 3 | Area 4 | Area 5 | Area 6 | Area 7 | Area 8 | Area 9 | Area 10 |
|---|---|---|---|---|---|---|---|---|---|---|
| $\mu_k$ | 2.806 | 4.072 | 3.141 | 0.638 | 0.398 | 1.223 | 2.321 | 2.212 | 1.078 | 2.106 |
| $\sigma_k$ | 1.216 | 1.052 | 1.211 | 1.569 | 1.300 | 1.599 | 1.198 | 0.988 | 1.145 | 1.253 |

| Parameter | Area 11 | Area 12 | Area 13 | Area 14 | Area 15 | Area 16 | Area 17 | Area 18 | Area 19 |
|---|---|---|---|---|---|---|---|---|---|
| $\mu_k$ | –0.323 | 0.382 | 3.020 | 1.749 | 3.041 | 1.550 | 3.070 | 1.244 | 0.938 |
| $\sigma_k$ | 1.088 | 1.335 | 0.803 | 1.003 | 1.122 | 1.477 | 0.962 | 0.858 | 1.214 |

As is to be expected, insurance losses in locally adjacent areas show a high degree of stochastic dependence, which can also be seen from the following correlation tables. For a better readability, only two decimal places are displayed.

**Table 6.** Empirical correlations between original losses in adjacent areas

|     | A1 | A2 | A3 | A4 | A5 | A6 | A7 | A8 | A9 | A10 | A11 | A12 | A13 | A14 | A15 | A16 | A17 | A18 | A19 |
|---|---|---|---|---|---|---|---|---|---|---|---|---|---|---|---|---|---|---|---|
| A1 | 1 | 0.46 | 0.03 | 0.16 | 0.47 | 0.20 | 0.35 | 0.49 | 0.41 | 0.24 | 0.78 | 0.64 | 0.91 | 0.63 | 0.85 | 0.66 | 0.30 | 0.67 | 0.56 |
| A2 | 0.46 | 1 | 0.64 | 0.78 | 0.67 | 0.36 | 0.51 | 0.76 | 0.57 | 0.51 | 0.58 | -0.04 | 0.59 | 0.84 | 0.68 | 0.58 | 0.87 | 0.77 | 0.90 |
| A3 | 0.03 | 0.64 | 1 | 0.93 | 0.41 | 0.26 | 0.11 | 0.16 | 0.33 | 0.16 | 0.08 | -0.09 | 0.13 | 0.64 | 0.25 | 0.10 | 0.74 | 0.14 | 0.35 |
| A4 | 0.16 | 0.78 | 0.93 | 1 | 0.54 | 0.36 | 0.16 | 0.25 | 0.43 | 0.19 | 0.22 | -0.10 | 0.30 | 0.79 | 0.36 | 0.19 | 0.84 | 0.32 | 0.49 |
| A5 | 0.47 | 0.67 | 0.41 | 0.54 | 1 | 0.41 | 0.35 | 0.51 | 0.84 | 0.63 | 0.59 | 0.02 | 0.64 | 0.67 | 0.59 | 0.50 | 0.58 | 0.71 | 0.67 |
| A6 | 0.20 | 0.36 | 0.26 | 0.36 | 0.41 | 1 | 0.07 | 0.11 | 0.28 | 0.19 | 0.28 | 0.14 | 0.31 | 0.42 | 0.24 | 0.27 | 0.39 | 0.27 | 0.40 |
| A7 | 0.35 | 0.51 | 0.11 | 0.16 | 0.35 | 0.07 | 1 | 0.44 | 0.27 | 0.19 | 0.48 | -0.07 | 0.46 | 0.35 | 0.45 | 0.91 | 0.64 | 0.61 | 0.49 |
| A8 | 0.49 | 0.76 | 0.16 | 0.25 | 0.51 | 0.11 | 0.44 | 1 | 0.50 | 0.75 | 0.61 | -0.03 | 0.54 | 0.47 | 0.71 | 0.53 | 0.40 | 0.75 | 0.90 |
| A9 | 0.41 | 0.57 | 0.33 | 0.43 | 0.84 | 0.28 | 0.27 | 0.50 | 1 | 0.66 | 0.68 | -0.01 | 0.52 | 0.60 | 0.50 | 0.41 | 0.46 | 0.65 | 0.63 |
| A10 | 0.24 | 0.51 | 0.16 | 0.19 | 0.63 | 0.19 | 0.19 | 0.75 | 0.66 | 1 | 0.33 | -0.12 | 0.27 | 0.28 | 0.43 | 0.24 | 0.23 | 0.45 | 0.65 |
| A11 | 0.78 | 0.58 | 0.08 | 0.22 | 0.59 | 0.28 | 0.48 | 0.61 | 0.68 | 0.33 | 1 | 0.19 | 0.79 | 0.65 | 0.80 | 0.73 | 0.43 | 0.84 | 0.74 |
| A12 | 0.64 | -0.04 | -0.09 | -0.10 | 0.02 | 0.14 | -0.07 | -0.03 | -0.01 | -0.12 | 0.19 | 1 | 0.44 | 0.21 | 0.28 | 0.17 | -0.12 | 0.13 | 0.03 |
| A13 | 0.91 | 0.59 | 0.13 | 0.30 | 0.64 | 0.31 | 0.46 | 0.54 | 0.52 | 0.27 | 0.79 | 0.44 | 1 | 0.71 | 0.86 | 0.74 | 0.47 | 0.76 | 0.65 |
| A14 | 0.63 | 0.84 | 0.64 | 0.79 | 0.67 | 0.42 | 0.35 | 0.47 | 0.60 | 0.28 | 0.65 | 0.21 | 0.71 | 1 | 0.74 | 0.54 | 0.79 | 0.68 | 0.72 |
| A15 | 0.85 | 0.68 | 0.25 | 0.36 | 0.59 | 0.24 | 0.45 | 0.71 | 0.50 | 0.43 | 0.80 | 0.28 | 0.86 | 0.74 | 1 | 0.69 | 0.47 | 0.71 | 0.75 |
| A16 | 0.66 | 0.58 | 0.10 | 0.19 | 0.50 | 0.27 | 0.91 | 0.53 | 0.41 | 0.24 | 0.73 | 0.17 | 0.74 | 0.54 | 0.69 | 1 | 0.63 | 0.77 | 0.64 |
| A17 | 0.30 | 0.87 | 0.74 | 0.84 | 0.58 | 0.39 | 0.64 | 0.40 | 0.46 | 0.23 | 0.43 | -0.12 | 0.47 | 0.79 | 0.47 | 0.63 | 1 | 0.59 | 0.64 |
| A18 | 0.67 | 0.77 | 0.14 | 0.32 | 0.71 | 0.27 | 0.61 | 0.75 | 0.65 | 0.45 | 0.84 | 0.13 | 0.76 | 0.68 | 0.71 | 0.77 | 0.59 | 1 | 0.86 |
| A19 | 0.56 | 0.90 | 0.35 | 0.49 | 0.67 | 0.40 | 0.49 | 0.90 | 0.63 | 0.65 | 0.74 | 0.03 | 0.65 | 0.72 | 0.75 | 0.64 | 0.64 | 0.86 | 1 |

**Table 7.** Empirical correlations between log losses in adjacent areas

|     | A1 | A2 | A3 | A4 | A5 | A6 | A7 | A8 | A9 | A10 | A11 | A12 | A13 | A14 | A15 | A16 | A17 | A18 | A19 |
|---|---|---|---|---|---|---|---|---|---|---|---|---|---|---|---|---|---|---|---|
| A1 | 1 | 0.27 | 0.30 | 0.16 | 0.17 | 0.45 | 0.28 | 0.32 | 0.32 | 0.29 | 0.67 | 0.51 | 0.76 | 0.34 | 0.67 | 0.74 | 0.18 | 0.21 | 0.29 |
| A2 | 0.27 | 1 | 0.48 | 0.66 | 0.39 | 0.37 | 0.71 | 0.69 | 0.52 | 0.64 | 0.30 | -0.02 | 0.45 | 0.66 | 0.58 | 0.45 | 0.73 | 0.74 | 0.78 |
| A3 | 0.30 | 0.48 | 1 | 0.70 | 0.40 | 0.31 | 0.42 | 0.51 | 0.58 | 0.53 | 0.18 | 0.07 | 0.21 | 0.32 | 0.54 | 0.26 | 0.47 | 0.21 | 0.57 |
| A4 | 0.16 | 0.66 | 0.70 | 1 | 0.77 | 0.47 | 0.46 | 0.47 | 0.59 | 0.49 | 0.18 | -0.13 | 0.33 | 0.50 | 0.47 | 0.18 | 0.76 | 0.43 | 0.54 |
| A5 | 0.17 | 0.39 | 0.40 | 0.77 | 1 | 0.59 | 0.30 | 0.20 | 0.49 | 0.39 | 0.28 | 0.08 | 0.35 | 0.56 | 0.44 | 0.16 | 0.55 | 0.36 | 0.41 |
| A6 | 0.45 | 0.37 | 0.31 | 0.47 | 0.59 | 1 | 0.14 | 0.01 | 0.36 | 0.34 | 0.33 | 0.12 | 0.48 | 0.46 | 0.48 | 0.37 | 0.59 | 0.17 | 0.50 |
| A7 | 0.28 | 0.71 | 0.42 | 0.46 | 0.30 | 0.14 | 1 | 0.52 | 0.27 | 0.40 | 0.45 | -0.07 | 0.31 | 0.31 | 0.46 | 0.62 | 0.63 | 0.58 | 0.57 |
| A8 | 0.32 | 0.69 | 0.51 | 0.47 | 0.20 | 0.01 | 0.52 | 1 | 0.64 | 0.81 | 0.27 | -0.02 | 0.38 | 0.35 | 0.56 | 0.35 | 0.28 | 0.62 | 0.63 |
| A9 | 0.32 | 0.52 | 0.58 | 0.59 | 0.49 | 0.36 | 0.27 | 0.64 | 1 | 0.78 | 0.40 | 0.19 | 0.27 | 0.50 | 0.44 | 0.30 | 0.33 | 0.57 | 0.61 |
| A10 | 0.29 | 0.64 | 0.53 | 0.49 | 0.39 | 0.34 | 0.40 | 0.81 | 0.78 | 1 | 0.21 | -0.02 | 0.21 | 0.37 | 0.52 | 0.30 | 0.31 | 0.53 | 0.81 |
| A11 | 0.67 | 0.30 | 0.18 | 0.18 | 0.28 | 0.33 | 0.45 | 0.27 | 0.40 | 0.21 | 1 | 0.47 | 0.49 | 0.45 | 0.60 | 0.67 | 0.20 | 0.45 | 0.39 |
| A12 | 0.51 | -0.02 | 0.07 | -0.13 | 0.08 | 0.12 | -0.07 | -0.02 | 0.19 | -0.02 | 0.47 | 1 | 0.44 | 0.21 | 0.24 | 0.46 | -0.23 | 0.25 | 0.05 |
| A13 | 0.76 | 0.45 | 0.21 | 0.33 | 0.35 | 0.48 | 0.31 | 0.38 | 0.27 | 0.21 | 0.49 | 0.44 | 1 | 0.55 | 0.60 | 0.71 | 0.37 | 0.39 | 0.24 |
| A14 | 0.34 | 0.66 | 0.32 | 0.50 | 0.56 | 0.46 | 0.31 | 0.35 | 0.50 | 0.37 | 0.45 | 0.21 | 0.55 | 1 | 0.59 | 0.43 | 0.57 | 0.58 | 0.53 |
| A15 | 0.67 | 0.58 | 0.54 | 0.47 | 0.44 | 0.48 | 0.46 | 0.56 | 0.44 | 0.52 | 0.60 | 0.24 | 0.60 | 0.59 | 1 | 0.59 | 0.36 | 0.35 | 0.63 |
| A16 | 0.74 | 0.45 | 0.26 | 0.18 | 0.16 | 0.37 | 0.62 | 0.35 | 0.30 | 0.30 | 0.67 | 0.46 | 0.71 | 0.43 | 0.59 | 1 | 0.38 | 0.43 | 0.39 |
| A17 | 0.18 | 0.73 | 0.47 | 0.76 | 0.55 | 0.59 | 0.63 | 0.28 | 0.33 | 0.31 | 0.20 | -0.23 | 0.37 | 0.57 | 0.36 | 0.38 | 1 | 0.52 | 0.56 |
| A18 | 0.21 | 0.74 | 0.21 | 0.43 | 0.36 | 0.17 | 0.58 | 0.62 | 0.57 | 0.53 | 0.45 | 0.25 | 0.39 | 0.58 | 0.35 | 0.43 | 0.52 | 1 | 0.60 |
| A19 | 0.29 | 0.78 | 0.57 | 0.54 | 0.41 | 0.50 | 0.57 | 0.63 | 0.61 | 0.81 | 0.39 | 0.05 | 0.24 | 0.53 | 0.63 | 0.39 | 0.56 | 0.60 | 1 |

The following graph shows estimated cdf's on a basis of 100,000 Monte Carlo simulations for the aggregated loss using lognormal margins with the parameters from Table 5 with a Bernstein copula representing **U** and a minimal correlation Gaussian copula representing **V**, for various values of $p$. For comparison purposes, we have also added an estimated cdf for the aggregated loss for a Bernstein copula representing **U** and an upper Fréchet (or comonotonicity) copula representing **V**. Note that the Bernstein copula is here constructed according to Cottin and Pfeifer (2014) on the basis of the ranks of the risk vectors, see also Pfeifer and Ragulina (2020), Chapter 3.

The following graphs for the tail cdf's correspond to a Bernstein copula **U** with a minimal correlation Gaussian copula **V**: $p=1$ $[F_1(x)]$; $p=0.99$ $[F_2(x)]$; $p=0.994$ $[F_3(x)]$ and a Bernstein copula **U** with $p=0.994$ but different copulas **V**: upper Fréchet bound or comonotonicity copula $[F_4(x)]$; independence copula $[F_5(x)]$.

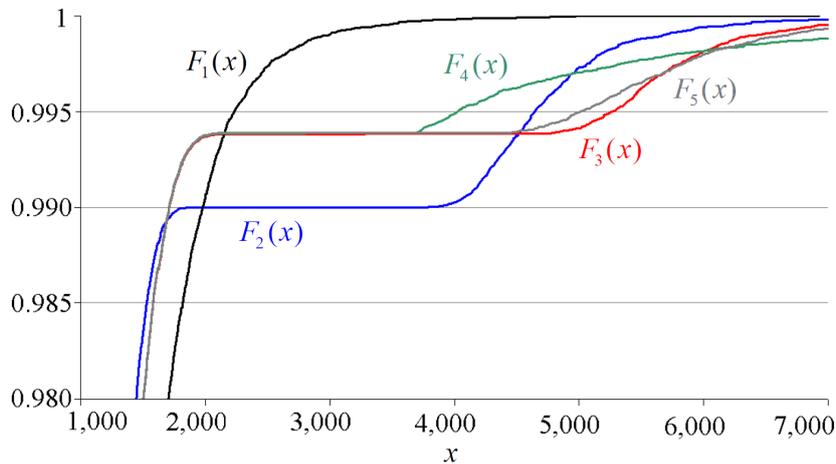

**Figure 15.** Plots of estimated cdf's $F_i(x)$, $i=1,\cdots,5$ in the tail

The following table shows the estimated risk measures $\text{VaR}_\alpha$ for $\alpha=0.005$ (Solvency II-standard) for the various values of $p$ and different types of **V**.

**Table 8.** Survey over VaR-estimates under different copula models

| $p$ | 0.990 | 0.994 | 0.994 | 0.994 | 1 |
|---|---|---|---|---|---|
| **V** | min corr Gauss | min corr Gauss | upper Fréchet | independence | --- |
| $\text{VaR}_\alpha$ | 4,647 MMU | 5,272 MMU | 3,976 MMU | 5,018 MMU | 2,229 MMU |

As can clearly be seen, the patchwork construction with the minimal correlation Gaussian copula representing **V** with no tail dependence gives the largest VaR estimate here and is typically larger than the construction with the upper Fréchet bound which has a positive tail dependence. Note that the sum of individual VaR's is given by 3,976 MMU which means that using the Bernstein copula alone would lead to a diversified portfolio while all other copula models do not.

Finally, it should be pointed out that the effects described here are independent of the particular copula chosen for **U**, i.e. the magnitude of the estimated VaR's under the patchwork construction would remain roughly equal also under an elliptical, an Archimedean, a vine or an adapted Bernstein copula approach for **U** (Pfeifer and Ragulina (2020)), cf. also the comments after Fig. 3 in Embrechts et al. (2013).

**4. Concluding remarks.**

The patchwork copula construction presented in this paper allows for a simple but yet effective and well-defined way to generate unfavourable VaR scenarios, i.e. scenarios with opposite diversification effects in particular for applications in Solvency II. Such scenario considerations are prescribed by legislative guidelines as e.g. specified in the Commission Delegated Regulation of the EU (2015), p. L12/6 (16), L12/9 (49), L12/12 (75) or (77), just to mention some. Besides Solvency II, such unfavourable VaR scenario generations could also be of interest in the Basel III framework (e.g. economic scenario generators) or in the reinsurance industry, in particular w.r.t. extreme natural perils.

Although there is theoretically also a method to create worst VaR scenarios by means of the rearrangement algorithm, the latter approach easily becomes numerically cumbersome in high-dimensional portfolios as in our case study, especially, if the risk distributions are not identical, see Embrechts et al. (2013), section 2.2. Hence a sub-optimal, but easy to implement alternative is of value, in particular, since it seems unlikely that the worst VaR scenario would actually occur in real life portfolios.

The approach discussed in this paper seems, at a first glance, to be related to the recent paper by Pfeifer and Ragulina (2018). The essential difference is, however, that the latter paper is not based on an observation-free copula construction for the tails as in the present paper. The algorithm proposed there leads only to stochastic approximations of the underlying distributions by a marginal-wise backwards transformation of the simulated multivariate distribution with the quantile functions of the originally estimated marginal cdf's. This emphasizes the fact that unfavourable VaR estimates cannot perhaps be characterized by the copula structure alone but that the interplay between the dependence structure and the marginal distributions is also essential, see the discussion in Ibragimov and Prokhorov (2017). Such a kind of interplay could potentially also be considered in the present approach, allowing non-constant negative pairwise correlations in the matrix $\Sigma_d$ for the Gaussian copula in Lemma 5.

Note that Value at Risk is not the only risk measure that is used for calculating capital requirements in Europe. For instance, the Swiss Solvency Test uses the Expected Shortfall $ES_\alpha(X)$ of risks $X$ (ES) as the underlying risk measure, cf. the Federal Office of Private Insurance, FINMA, Switzerland (2006). In accordance with our terminology and under the assumption of a continuous risk distribution, it is defined as

$$ES_\alpha(X) = E(X | X > VaR_\alpha(X)), \quad 0 < \alpha < 1.$$

Unfortunately it is impossible to generate true unfavourable ES scenarios since ES is a coherent (i.e. subadditive) risk measure which in the worst case generates additive risk scenarios if the risks involved follow a comonotone dependence structure (see e.g. McNeill et al. (2015), Chapter 7.2). Note however, that it is sufficient for the generation of additive ES scenarios to use a dependence structure as in Lemma 1 with the upper Fréchet bound for **V** which is a copula in any dimension.

**Appendix: Proof of Lemmata.**

**Proof of Lemma 2.** We have

$$F_{Z_{1i}}(x,\beta) = P(Q_i((1-\beta) \cdot U_i) \leq x) = P((1-\beta) \cdot U_i \leq F_i(x)) = P\left(U_i \leq \frac{F_i(x)}{1-\beta}\right) = \frac{F_i(x)}{1-\beta}$$

for $0 \leq x \leq Q_i(1-\beta)$ and

$$F_{Z_{2i}}(x,\beta) = P(Q_i(1-\beta+\beta \cdot V_i) \leq x) = P(1-\beta+\beta \cdot V_i \leq F_i(x)) = \frac{F_i(x)+\beta-1}{\beta}$$

for $x \geq Q_i(1-\beta)$, $i = 1, 2$. ●

**Proof of Lemma 3.** In the finite interval case, we have, by the usual convolution formula,

$$h_1(x) = \int_{\substack{0 \leq y \leq M \\ 0 \leq x-y \leq M}} f(x-y)g(y)\,dy = \int_{\max(0,x-M) \leq y \leq \min(x,M)} f(x-y)g(y)\,dy.$$

Now for $0 \leq x \leq M$, we have $\max(0, x-M) = 0$, $\min(x, M) = x$, from which the upper formula in brackets for $h_1(x)$ follows. For $M \leq x \leq 2M$, we have $\max(0, x-M) = x-M$, $\min(x, M) = M$, from which the lower formula in brackets for $h_1(x)$ follows.

The proof for the infinite interval case is analogous, observing that for $x \geq 2M$, we have

$$h_2(x) = \int_{\substack{M \leq y \leq x \\ M \leq x-y}} f(x-y)g(y)\,dy = \int_{M \leq y \leq x-M} f(x-y)g(y)\,dy.$$

Further, under the conditions made, we have, in either case,

$$\frac{d}{dx}F*G(x)\bigg|_{x=2M} = h_1(2M) = h_2(2M) = \int_M^M f(x-y)g(y)\,dy = 0,$$

as stated. ●

**Proof of Lemma 4.** Let $\xi_i$ and $\zeta_i$ be independent random variables with the cdf's $\underline{F}(\bullet,\beta)$ and $\overline{F}(\bullet,\beta)$, resp. Then $I \cdot \xi_i + (1-I)\cdot(Q(1-\beta)+\zeta_i)$ is a stochastic representation of $X_i$, $i=1,\cdots,d$, where again $I$ is a binomial random variable with $P(I=1)=1-\beta$ and $P(I=0)=\beta$, independent of $(\mathbf{U},\mathbf{V})$, according to Lemma 2. Hence

$$I\cdot\sum_{i=1}^{d}\xi_i + (1-I)\cdot\sum_{i=1}^{d}(Q(1-\beta)+\zeta_i) = I\cdot\sum_{i=1}^{d}\xi_i + (1-I)\cdot\left[dQ(1-\beta)+\sum_{i=1}^{d}\zeta_i\right]$$

is a stochastic representation of $S$. Note that the cdf of $\sum_{i=1}^{d}\xi_i$ is $\underline{F}^{d*}(\bullet,\beta)$ and that of $\sum_{i=1}^{d}\zeta_i$ is $\overline{F}^{d*}(\bullet,\beta)$, from which the assertion follows. ●

**Proof of Lemma 5.** The proof relies on the following two relations:

a) $\sum_{k=2}^{d}\frac{1}{k(k-1)} = \frac{d-1}{d}$ for $d \geq 2$,

b) $\frac{i-1}{i} + \sum_{k=1}^{d-i}\frac{1}{(i+k)(i+k-1)} = \frac{d-1}{d}$ for $d \geq 2$, $1 \leq i \leq d$.

Clearly a) follows easily by induction. Relation b) follows immediately from a) since

$$\frac{i-1}{i} = \sum_{k=2}^{i}\frac{1}{k(k-1)} \text{ and } \sum_{k=1}^{d-i}\frac{1}{(i+k)(i+k-1)} = \sum_{k=i+1}^{d}\frac{1}{k(k-1)}.$$

To prove Lemma 5, we first show that $\mathbf{T}\mathbf{T}^{tr} = \mathbf{I}_d = \mathbf{T}^{tr}\mathbf{T}$. Let $\mathbf{T}\mathbf{T}^{tr} = [b_{ij}]_{i,j=1,\cdots,d}$. For $1 \leq i \leq d$ we obtain, by relation b) above, $b_{ii} = \frac{1}{d} + \frac{i-1}{i} + \sum_{k=1}^{d-i}\frac{1}{(i+k)(i+k-1)} = 1$. For $1 \leq i,j \leq d$ with $i \neq j$ we get, with $i \vee j := \max(i,j)$, following again relation b),

$$b_{ij} = \frac{1}{d} - \frac{1}{i\vee j} + \sum_{k=i\vee j+1}^{d}\frac{1}{k(k-1)} = \frac{1}{d} - \frac{1}{i\vee j} + \sum_{k=1}^{d-i\vee j}\frac{1}{(k+i\vee j)(k+i\vee j-1)}$$

$$= \frac{1}{d} - \frac{1}{i\vee j} + \frac{d-1}{d} - \frac{i\vee j-1}{i\vee j} = 1-1 = 0.$$

This proves $\mathbf{T}\mathbf{T}^{tr} = \mathbf{I}_d$. On the other hand, let $\mathbf{T}^{tr}\mathbf{T} = [c_{ij}]_{i,j=1,\cdots,d}$. It is obvious that

$c_{11} = \frac{1}{d} \cdot d = 1$ and for all $2 \leq i \leq d$, $c_{ii} = \frac{1}{i(i-1)} \cdot (i-1) + \frac{i-1}{i} = 1$.

Next, for all $2 \leq j \leq d$, we obtain $c_{1j} = \frac{1}{\sqrt{d}}\left(-\frac{1}{\sqrt{j(j-1)}} \cdot (j-1) + \sqrt{\frac{j-1}{j}}\right) = 0$, and for all

$2 \leq i \leq d$, we get $c_{i1} = \frac{1}{\sqrt{d}}\left(-\frac{1}{\sqrt{i(i-1)}} \cdot (i-1) + \sqrt{\frac{i-1}{i}}\right) = 0$. Finally, for $2 \leq i, j \leq d$ with

$i \neq j$, we get $c_{ij} = -\frac{1}{\sqrt{(i \vee j) \cdot (i \vee j - 1)}} \cdot \left(-\frac{1}{\sqrt{(i \vee j) \cdot (i \vee j - 1)}} \cdot (i \vee j - 1) + \sqrt{\frac{i \vee j - 1}{i \vee j}}\right) = 0$.

This proves $\mathbf{T}^{tr}\mathbf{T} = \mathbf{I}_d$. Now let $\lambda_1 = 1 + (d-1)r$, $\lambda_i = 1 - r$, $i = 2,\cdots,d$ and $\Delta_t = \mathrm{diag}(\lambda_1 - t, \cdots, \lambda_d - t)$. A standard computation yields, for $t \in \mathbb{R}$, $\mathbf{T}\Delta_t = [s_{ij}]_{i,j=1,\cdots,d}$ where

$$s_{ij} = \begin{cases} \dfrac{1+(d-1)r-t}{\sqrt{d}}, & j=1 \\ -\dfrac{1-r-t}{\sqrt{j(j-1)}}, & 1 \leq i < j \\ \sqrt{\dfrac{j-1}{j}}(1-r-t), & 1 < i = j \\ 0, & \text{otherwise.} \end{cases}$$

Let $\mathbf{T}\Delta_t\mathbf{T}^{tr} = [d_{ij}]_{i,j=1,\cdots,d}$. From relation a) above it follows that

$d_{11} = \frac{1+(d-1)r-t}{d} + (1-r-t)\sum_{k=2}^{d}\frac{1}{k(k-1)} = \frac{1+(d-1)r-t}{d} + (1-r-t) \cdot \frac{d-1}{d} = 1-t$,

and for $2 \leq i \leq d$, relation b) gives

$d_{ii} = \frac{1+(d-1)r-t}{d} + (1-r-t)\left(\frac{i-1}{i} + \sum_{k=1}^{d-i}\frac{1}{(i+k)(i+k-1)}\right)$

$= \frac{1+(d-1)r-t}{d} + (1-r-t)\frac{d-1}{d} = 1-t$.

Next, for $2 \leq i, j \leq d$ with $i \neq j$ we obtain from relation b) above,

$d_{ij} = \frac{1+(d-1)r-t}{d} - \frac{1-r-t}{i \vee j} + (1-r-t)\left(\sum_{k=1}^{d-i \vee j}\frac{1}{(i \vee j + k)(i \vee j + k - 1)}\right)$

$= \frac{1+(d-1)r-t}{d} - \frac{1-r-t}{i \vee j} + (1-r-t)\left(\frac{d-1}{d} - \frac{i \vee j - 1}{i \vee j}\right) = r$.

This in turn means $\mathbf{T}\Delta_t\mathbf{T}^{tr} = \Sigma_d - t\mathbf{I}_d$. Consequently, the characteristic polynomial for $\Sigma_d$ is given by

$$\varphi_{\Sigma_d}(t) = \det(\Sigma_d - t\mathbf{I}_d) = \det(\mathbf{T}\Delta_t\mathbf{T}^{tr}) = \det(\mathbf{T})\cdot\det(\Delta_t)\cdot\det(\mathbf{T}^{tr}) = \det(\mathbf{T})\cdot\det(\Delta_t)\cdot\det(\mathbf{T}^{-1})$$

$$= \det(\Delta_t) = \prod_{i=1}^{d}(\lambda_i - t).$$

Hence $\lambda_i$, $1 \leq i \leq d$, are the latent roots of $\Sigma_d$. Therefore, $\Sigma_d$ is a correlation matrix, i.e. positive semidefinite iff $\lambda_i \geq 0$ for all $1 \leq i \leq d$, i.e. $-\dfrac{1}{d-1} \leq r \leq 1$.

Thus Lemma 5 is proved. ●

**Acknowledgements.** We would like to thank the referees for several helpful comments that improved the presentation of the paper essentially.